\documentclass[letterpaper,twocolumn,10pt]{article}
\usepackage{graphics}
\usepackage{endnotes}
\usepackage{usenix}
\usepackage{xspace}
\usepackage{url}
\usepackage{verbatim}
\usepackage{amssymb}

\newcommand{\eat}[1]{}

\newcommand{\MTTF}{\ensuremath{\mathrm{MTTF}}\xspace}

\newcommand{\ML}{\ensuremath{\mathrm{ML}}\xspace}
\newcommand{\MV}{\ensuremath{\mathrm{MV}}\xspace}
\newcommand{\MRL}{\ensuremath{\mathrm{MRL}}\xspace}
\newcommand{\MRV}{\ensuremath{\mathrm{MRV}}\xspace}
\newcommand{\MD}{\ensuremath{\mathrm{MDL}}\xspace}

\newcommand{\MTTDL}{\ensuremath{\mathrm{MTTDL}}\xspace}

\newenvironment{MYitemize}{\begin{list}{\labelitemi}{%
\setlength{\topsep}{0pt plus 0pt minus 0pt}%
\setlength{\itemsep}{0pt plus 0pt minus 0pt}%
\setlength{\parsep}{0pt plus 0pt minus 0pt}%
\setlength{\parskip}{0pt plus 0pt minus 0pt}%
\setlength{\leftmargin}{\leftmarginiv}%
\setlength{\listparindent}{1em}%
}}{\end{list}}

\newcounter{MYenumctr}

\headheight \baselineskip
\begin{document}
\ifx\href\undefined\else\hypersetup{linktocpage=true}\fi 

\date{}
\title{\Large \bf A Fresh Look at the Reliability of Long-term Digital Storage}

\author{
{\rm Mary Baker}\\
 HP Labs, Palo Alto, {CA}
\and
{\rm Mehul Shah}\\
 HP Labs, Palo Alto, {CA}
\and
{\rm David S. H. Rosenthal}\\
 Stanford University Libraries, {CA}
\and
{\rm Mema Roussopoulos}\\
 Harvard University, Cambridge, {MA}
\and
{\rm Petros Maniatis}\\
 Intel Research, Berkeley, {CA}
\and
{\rm TJ Giuli}\\
 Stanford University, {CA}
\and
{\rm Prashanth Bungale}\\
 Harvard University, Cambridge, {MA}
}

\maketitle

\thispagestyle{empty}

\subsection*{Abstract}

Many emerging Web services, such as email, photo sharing, and web site
archives, need to preserve large amounts of quickly-accessible data
indefinitely into the future. In this paper, we make the case that
these applications' demands on large scale storage systems over long
time horizons require us to re-evaluate traditional storage system
designs. We examine threats to long-lived data from an end-to-end
perspective, taking into account not just hardware and software faults
but also faults due to humans and organizations. We present a simple
model of long-term storage failures that helps us reason about the
various strategies for addressing these threats in a cost-effective
manner. Using this model we show that the most important strategies
for increasing the reliability of long-term storage are detecting
latent faults quickly, automating fault repair to make it faster and
cheaper, and increasing the independence of data replicas.

\section{Introduction}
\label{sec:introduction}

Frequent headlines remind us that bits, even bits stored in expensive,
professionally administered data centers, are vulnerable to loss and
damage~\cite{Hansen2004,ISAICAI2004,Leyden2002,stanfordGSBBackup}.
Despite this, emerging web services such as e-mail (e.g., Gmail),
photo sharing (e.g., Snapfish, Ofoto, Shutterfly)\footnote{All
trademarks mentioned in this paper belong to their respective owners.}
and archives (e.g., The Internet Archive) require
large volumes of data to be stored indefinitely into the future.  The
economic viability of these services depends on storing the data at
very low cost.  Customer acceptance of these services depends on the
data remaining both unaltered and accessible with low latency.

Satisfying these requirements over long periods of time would be easy
if fast, cheap, reliable disks were available, and if threats to the data
were confined to the storage subsystem itself.  Unfortunately, neither is
true.  The economics
of high-volume manufacturing provide a choice between consumer-grade drives,
which are cheap, fairly fast, and fairly reliable,
and enterprise-grade drives, which are vastly more expensive, much faster
but only a little more reliable.  In \S~\ref{sec:strategies:MTTFV} we show
that an enterprise drive fourteen times more expensive
than a consumer drive only reduces bit errors from 8 to 6 over
a 5-year, 99\% idle lifetime.  For short-lived data, this level of
reliability might not pose a problem, but for long-lived data these
faults are inescapable.

Further, long-term storage faces many threats beyond the
storage system itself.  These include
obsolescence of data formats, long-term attacks on the data, and economic
and structural volatility of organizations sponsoring the storage.

In this paper, we make the case that long-term reliable storage is
a problem that deserves a fresh look, given its significant differences
from traditional storage problems addressed in the literature.  We
start by motivating the need for \emph{digital preservation} --- storing
immutable data over
long periods (\S~\ref{sec:motivation}).  We then list the threats that
imperil data survival using examples from real systems
(\S~\ref{sec:threats}), and examine why the design philosophy of many
current storage systems is insufficient for long-term storage
(\S~\ref{sec:dangerous}).

To understand the implications of this problem better, we introduce a
simple reliability model (\S~\ref{sec:analyticModel2}) of replicated
storage systems designed to address long-term storage threats.  This
model is inspired by the reliability model for
RAID~\cite{Patterson1988short}, but our extensions and
interpretation of the model take a more holistic, end-to-end, rather
than device-oriented, approach. Our model includes a wider range of 
faults to which the replicas are subject. It explicitly
incorporates both \emph{latent faults}, which occur long before they
are detected, and \emph{correlated faults}, when
one fault causes others or when multiple faults result from
the same error (\S~\ref{sec:dangerous}).  Our model
incorporates and highlights the importance of a detection process for
latent faults.

Although our model is simplistic, it highlights needed areas for
gathering reliability data, and it helps evaluate strategies for
improving the reliability of systems for long-lived data
(\S~\ref{sec:strategies}).  For example, we would like to be able to
answer questions such as, how dangerous are latent faults over long
time periods?  (Quite dangerous,
\S~\ref{sec:analyticModel2:implications}.)  Would it be better to
replicate an archive on tape or on disk?  (Disk,
\S~\ref{sec:strategies:MD}.)  Is it better to increase the mean time
between visible faults or between latent faults? (Perhaps neither if
it significantly decreases the other,
\S~\ref{sec:analyticModel2:implications}.)  Is it better to increase
replication in the system or increase the independence of existing
replicas?  (Both, but replication without increasing independence does
not help much,
\S~\ref{sec:analyticModel2:replication}.)
Some of these strategies have been proposed before
(\S~\ref{sec:related}), but we believe they are worth revisiting
in the context of long-term storage.

We conclude (\S~\ref{sec:conclusion}) that the
most important strategies for increasing reliability of long-term storage
are detecting latent faults quickly, automating the repair of faults to
be faster and cheaper, and increasing the
independence of data replicas.  This
includes increasing geographic, administrative, organizational, and
third-party independence, as well as the diversity of hardware and
software.  We hope that our analysis and conclusions, while still
primitive, will motivate a
renewed look at the design of storage systems that must preserve data for
decades in volatile and even hostile environments.

\section{The Need for Long-term Preservation}
\label{sec:motivation}

Preserving information for decades or even centuries has proved beneficial
in many cases.
In the 12th century BC Shang dynasty Chinese astronomers inscribed eclipse
observations on ``oracle bones'' (animal bones and tortoise shells).
About 3200 years later researchers used these records, together with
one from 1302BC, to estimate that the total clock error
that had accumulated was just over 7 hours,  and from this derived
a value for the viscosity of the Earth's mantle as it rebounds from
the weight of the glaciers~\cite{Pang1995}.

Longitudinal studies in the field
of medical research
depend upon accurate preservation of detailed patient records for
decades.  In 1948 scientists began to study the residents of
Framingham, Massachusetts~\cite{Dawber1951}
to understand the large increase in heart
disease victims throughout the 1930s and 40s.  Using data collected
over decades of research, scientists discovered the major risk factors
that modern medicine now knows contribute to heart disease.

In 1975, the former USSR sent two probes, Venera 9 and 10, to the
surface of Venus to collect data and photographs.  The resulting
photographs were of very low quality and were relegated to the dustbin
of science history.  About 28 years later, an American scientist was able to
use
modern digital image processing algorithms to enhance the photographs to
reveal much more detailed images~\cite{Venus2003}.

These timescales of many decades, even centuries, contrast with
the typical 5-year lifetime for computing hardware and similar
lifetimes attributed to digital media.  It is not just scientific data
that is expected to persist over these timescales.  Legislation such
as Sarbanes-Oxley~\cite{Sarbanes-Oxley} and HIPPA~\cite{HIPPA} require
many companies and organizations to keep electronic records over
decades.  In addition, consumers used to analog assets such as mail
and photographs, which persist over many decades, are now happily
entrusting their digital versions to online services.  The associated marketing
literature~\cite{GmailArchive} encourages them to expect similar
longevity.

\section{Threats to Long-term Preservation}
\label{sec:threats}

While traditional short-term storage applications must also anticipate
a variety of threats, the list of threats to long-term storage
is longer and more diverse.  Some threats, such as media and software
obsolescence, are particular to long-term
preservation.  Other threats, such as  natural disaster and human
error, are also threats to
short-term storage applications, but the probability of their occurrence is
higher over the longer desired lifetimes for archival data.
In this section we list threats to long-term storage
and provide real examples to motivate them.
We take an end-to-end approach 
in identifying failures, concentrating not just on storage device faults but
also on faults in the environment, processes, and support surrounding the
storage systems.

\noindent{\bf Large-scale disaster.} Large-scale disasters, such as
floods,
fires, earthquakes, and acts of war must be anticipated
over the long desired lifetimes of archival data.  Such disasters
will typically be manifested by other types of threat, such as media,
hardware, and organizational faults, and sometimes all of these, as
was the case with many of the data centers affected by the 9/11 attack on the
World Trade Center.

\noindent{\bf Human error.}
One of the ways in which data are lost is through
users or operators accidentally deleting (or marking as overwriteable) content
they still need, or
accidentally or purposefully deleting data for which they later discover
a need.  For instance, rumors
suggest that large professionally administered digital repositories have
suffered administrative
errors that caused data loss across replicas,
but the organizations hosting these repositories are unwilling
to report the mishaps publicly (see \S~\ref{sec:conclusion}).
Sometimes the errors instead affect hardware (e.g.,
tapes being lost in transit~\cite{Reuters2005}), software (e.g., uninstalling a required driver),
or infrastructure (e.g., turning off the air-conditioning system in the
server room) on which the preservation application runs.  Human
error is increasingly the cause of system failures~\cite{Oppenheimer2003,Reason1990short}.

\noindent{\bf Component faults.}
Taking an end-to-end view of a system, all components can be expected to fail,
including hardware, software, network interfaces, and even third-party
services.
Hardware components
suffer transient recoverable faults (e.g., caused by
temporary power loss) and
catastrophic irrecoverable faults (e.g., a power surge fries a controller card).
Software components, including firmware in disks, suffer
from bugs that pose a risk to the stored data.
Systems cannot assume that the network transfers
they use to ingest or disseminate content
will terminate within a specified time period,
or will actually deliver the content unaltered. (The initial ingestion of large
collections into a repository is thus itself error-prone.)
Third-party components that cannot easily themselves be preserved are
also sources of problems.
External license servers or
the companies that run them might no longer exist decades after an application
and its data are archived.
Domain names
will vanish or be reassigned if the registrant fails to pay the registrar,
and a persistent URL~\cite{OCLCPURL} will not resolve if
the resolver service
fails to preserve its data with as much care as the storage system
client.

\noindent{\bf Media faults.}
The storage medium is a component of particular interest.  No
affordable digital storage medium is completely reliable over long
periods of time, since the medium can degrade; degradation and other
such medium errors can cause irrecoverable
bit faults, often called \emph{bit rot}.  Storage media are also subject
to sudden irrecoverable loss of bulk data from, for instance,
disk crashes~\cite{Talagala1999}.

Bit rot is particularly troublesome, because it occurs without warning and
might not be detected until it is too late to make repairs.
The most familiar example might be CD-ROMs.  Studies
of CD-ROMs~\cite{PCActive2003,Mald04} indicate that
despite being sold as reliable for decades, or even 75 to 100 years, they
are often only good for two to five years, even when stored in accordance with
the manufacturer's recommendations.

Disks are subject to similar problems.  For instance, a previously readable
sector can become unreadable.  Or a sector might be readable but contain
the wrong information due to a previously misplaced sector
write, often due to problems with vibrations~\cite{Anderson2003}.

\noindent{\bf Media/hardware obsolescence.}
Over time, media and hardware components
can become obsolete --- no
longer able to communicate with other system components --- or
irreplaceable after a fault.
This problem is particularly acute
for removable media, which have a long history of remaining theoretically
readable if only a suitable reader could be found~\cite{Keeton2001}.
Examples include 9-track tape and 12-inch video laser discs.
Evolution of the industry specification for PCs has made it
difficult to purchase a commodity PC with
built-in floppy drive, indicating that even floppy disks, once considered an
obvious ubiquitous cheap storage medium, are endangered.

\noindent{\bf Software/format obsolescence.}
Similarly, software components become
obsolete.  This is often manifested as \emph{format obsolescence}:
the bits in which the data were encoded remain accessible,
but the information can no longer be correctly interpreted.  Proprietary
formats, even those in widespread use,
are very vulnerable.
For instance, digital camera companies have their own proprietary ``RAW''
formats for recording the
raw data from their cameras.  These formats are often undocumented, and
when a company ceases to exist or to support its format, photographers can
lose vast amounts of data~\cite{openraw}.

\noindent{\bf Loss of context.}
Metadata,
or more generally ``context,'' includes information about layout, location,
and inter-relationships among stored objects, as well as the subject and
provenance of content, and the processes, algorithms and software needed
to manipulate them.  Preserving contextual metadata is as important
as preserving the actual data, and it can even be hard to recognize
all required context in time to collect it.  Encrypted information is
a particularly challenging
example, since preservation of the decryption keys is essential
alongside preservation of the encrypted data.  Unfortunately, over long
periods of time, secrets (such as decryption keys) tend to get lost, to leak,
or to get broken~\cite{diffieQuote}.  This problem is less of a threat to
short-term storage applications, where the assets do not live long enough
for the context to be lost or the information to become uninterpretable.

\noindent{\bf Attack.}
Traditional repositories are subject to long-term malicious attack, and
there is no reason to expect their digital equivalents to be exempt.
Attacks include
destruction, censorship, modification and theft of repositories' contents,
and disruption of their services~\cite{BookMutilation}.
The attacks can be short-term or long-term, legal or illegal, internal or
external. They can be motivated by ideological, political, financial
or legal factors, by bragging rights or by employee dissatisfaction.
Attacks are a threat to
short-term storage as well, but researchers usually focus on short-term,
intense attacks, rather than the slowly subversive attacks that
afflict long-term repositories.  Because much abuse of computer systems
involves insiders ~\cite{Keeney2005}, a digital preservation system must
anticipate attack even if it is completely isolated from external networks.
Examples of attacks include cases of ``sanitization'' of US government
websites to conform with the administration's world
view~\cite{departmentOfEducationWebScrubbing,whiteHouseWebScrubbing}. 

\noindent{\bf Organizational faults.}
A system view of long-term storage must
include not merely the technology but also the organization in which it is
embedded.  These organizations can die out, perhaps through bankruptcy, or
change missions.  This can deprive the storage technology
of the support it needs to survive.  System planning must envisage the
possibility of the asset represented by the preserved content being
transferred to a successor organization, or otherwise being properly
disposed of with a data ``exit strategy.''  Storage services can also
make mistakes, and assets dependent on a single service can be lost.

As an example, during an organizational change, a large IT company
closed a research lab and requested the lab's
research projects be copied to tape and sent to another of its labs.
Unfortunately, few knew about these tapes
and they were allowed to languish without documentation of their
contents.  When it became clear that some of the project data would be
useful to current researchers, enough time had passed that nobody could
identify what would be on which tape, and the volume of data was too huge
to reconstruct an index~\cite{Lau2004}.

As another example, a user lost all of her digital photos stored at
Ofoto when she did not make a purchase within the required interval and
did not receive their email warnings due to having failed to send them
her updated email address~\cite{Lazarus2005}.  Further, for many of
these services
there is no ``data exit strategy'' for easy bulk retrieval of the original
high-resolution format of all of of the customer's photos.

\noindent{\bf Economic faults.}
Many organizations with materials to
preserve do not have large budgets to apply to the problem, and
declare success after just managing to get a collection put online.
Unfortunately, this provides no plan for maintaining a collection's
accessibility or quality in the future.
There are ongoing costs
for power, cooling, bandwidth, system administration, equipment space,
domain registration, renewal of equipment, and so on.
Information in digital form is much more vulnerable
to interruptions in the money supply than information on paper, and
budgets for digital preservation must be expected to vary up and down,
possibly even to zero, over time.  These budget issues
affect our ability to preserve as many collections as desired: many
libraries now subscribe to fewer serials and monographs~\cite{ARLstats}.

Motivating an investment
in preservation can be difficult~\cite{Gray2002} without better
tools to predict long-term costs, especially
if the target audience for the preserved
information does not exist at the time decisions are made.
While budget is an issue
in the purchase of any storage system, it is usually easier to plan how 
money will be spent over a shorter lifespan.

\section{Dangerous Assumptions}
\label{sec:dangerous}

Some of the threats above are well-understood and we should have
succeeded in solving these problems by now.  So why do we still lose data?
Part of the reason is that our system designs often fail to take an
end-to-end perspective, and that, in practice,
we often make a few key, but potentially dangerous assumptions.
These include visibility
of faults (\S~\ref{sec:dangerous:visibility}), independence of faults (\S~\ref{sec:dangerous:independence}), and enough money
to apply exotic solutions (\S~\ref{sec:dangerous:money}), as described below.

\subsection{The fault visibility assumption}
\label{sec:dangerous:visibility}
While many faults are detected at the time an error causes them, some occur silently.
These are called ``latent faults.''  There are many sources of latent faults,
but media errors are the best known.
While a head crash would be detectable,
bit rot might not be uncovered until the
affected faulty data are actually accessed
and audited.  As another example, a sector on a disk might become unreadable;
this would not
be detected until the next read of that sector.
Further, a sector might be readable but contain
incorrect information due to a previous misplaced sector
write.  Silent media errors and faults occur more frequently than many
of us have assumed; for instance Schwarz {\em et al.} suggest that silent
block faults occur five times as often as whole disk faults~\cite{Schwarz2004}.

In aggregate, archives such as the Internet Archive might supply
users with data items at a high rate, but the average data item is accessed
infrequently.  Detecting loss or corruption only when a user requests
access renders the average data item vulnerable to an accumulation
of latent faults~\cite{KariThesis,Martin2005,Schwarz2004}.

While the need to guard against latent faults has long been
recognized in larger systems~\cite{Bartlett2004,KariThesis,Schwarz2004},
increases in storage capacity have recently brought more attention
to the problem for commodity storage.
An important example is the IRON File System~\cite{Prabhakaran2005},
which uses redundancy within a single disk to address
latent faults within file system metadata structures.

Beyond media faults, there are many types of latent faults caused
by threats in \S~\ref{sec:threats}:
\begin{MYitemize}
\item \emph{Human error}: accidental deletion or overwriting of materials might
not be discovered until those materials are needed.
\item \emph{Component failure}: The reliance on a failed system component or
a third-party component that
is no longer available might not be discovered until data depending on that
component are accessed.
\item \emph{Media/hardware obsolescence}: Failure of an obsolete, seldom-used
media reader might not be discovered until information on the associated medium
needs to be read.  It might then be impossible or too costly to
purchase another reader.
\item \emph{Software/format obsolescence}: Upon accessing old information
we might discover it is in a format belonging to an application we can no
longer run.
\item \emph{Loss of context}: We might not discover we are missing crucial metadata
about saved data until we try to make sense of the data.  For instance,
we might not have preserved an encryption key.
\item \emph{Attack}: Results of a successful censorship
or corruption attack on a
data repository might never be discovered, or might only become apparent upon
accessing the data long after the attack.
\end{MYitemize}

The general solution to latent faults is to detect them as quickly
as possible, as indicated by our model in \S~\ref{sec:strategies}.
For instance, ``scrubbing''~\cite{KariThesis,Maniatis2005short,Schwarz2004}
can be used to detect media faults
or evidence of data corruption from an attack.
If preserving the data is important, scrubbing should be performed while
the data can still be repaired from a replica (or error correction codes).
At a higher layer, we can prevent latent faults in the making by
detecting the need to migrate content from
old to new media or from old to new data formats before we have lost the
ability to read the old medium or interpret the old data format.  Another
example is detecting the need to re-encrypt old materials with new keys
before old keys are considered obsolete.

\subsection{The independence assumption}
\label{sec:dangerous:independence}

Many researchers and designers correctly point to
data replication as a strategy for preventing data loss, but make the
assumption that replicas fail independently.
Alas, in practice, faults are not as independent as we might hope.
Talagala~\cite{Talagala1999} logged every fault in a large disk farm
(of 368 disk drives) at UC Berkeley over six months and showed
significant correlations between them.
(For example, if power units are shared across drives, a single
power outage can affect a large number of drives.)
A single power outage accounted for 22\% of all machine
restarts.
Temperature and vibrations of tightly packed devices
in machine room racks~\cite{Anderson2003,Maltzahn2005} are also sources of
correlated media failures.

Taking an end-to-end perspective, there are many sources of fault
correlation corresponding to
the threats in \S~\ref{sec:threats}:

\begin{MYitemize}
\item \emph{Large-scale disaster}: A single large disaster might destroy all
replicas of the data.
Geographic replication clearly helps, but care
must be taken to ensure it provides sufficient independence. For
example, in the 2001 9/11 disaster in New York City, a data center was
destroyed.  The system failed over to a replicated data center on the
other side of a river, and the failover worked correctly.  Unfortunately,
the sites were still not sufficiently distant; the chaos in the streets
prevented staff from getting to the backup data center.
Eventually it was unable to continue unattended~\cite{Towers2004}.
\item \emph{Human error}: 
System administrators are human and fallible.
Unfortunately,
in most systems they are also very powerful,
able to destroy or modify data without restriction.
If all replicas are under unified administrative control, a single
human error can cause faults at all of them.
\item \emph{Component faults}: If all replicas of the information are
dependent on the same external component, for instance a license server,
the loss of that component causes correlated faults at every replica.
\item \emph{Loss of context}: Losing metadata associated with archival data might
cause correlated faults across replicas.  For instance, if materials are
encrypted with the same key at all replicas, loss of that key will make all
the replicas useless.
\item \emph{Attack}: Attacks can cause correlated faults.  For instance, a
flash worm might affect many replicas at
once.
\item \emph{Organizational faults}: Long-term storage must anticipate the
failure of any one organization or service.   Increasing the visibility
of digital assets and developing simple exit strategies for data is
important, but so is minimizing dependence on any single organization.
\end{MYitemize}

As indicated in \S~\ref{sec:strategies} by our model, the best way
to avoid correlated faults is independence of the replicas.  Simply
increasing the replication is not enough if we do not also ensure the
independence of the replicas geographically, administratively, and otherwise.

\subsection{The unlimited budget assumption}
\label{sec:dangerous:money}

The biggest threats to digital preservation are
economic faults.  With a lot of money we could apply many solutions to
preserving data, but
most of the information people would like to see live forever
is not in the hands of organizations with unlimited budgets.
Solutions such as synchronous mirroring of RAIDs across widely dispersed
geographic replicas might not be affordable over the long term for many
organizations.  Although we make a qualitative attempt to compare the costs
of some of the strategies and solutions we explore in this paper, 
estimation of these costs remains very difficult and is an area richly
deserving further work.

\section{Analytic Model}
\label{sec:analyticModel2}

Abstract models such as that in Patterson {\em et al.}~\cite{Patterson1988short} and Chen {\em et al.}\cite{Chen1994} are
useful for reasoning about the
reliability of different replicated storage system designs. In
this section, we build upon these models to further incorporate the
effect of latent and correlated faults on the overall reliability of
an arbitrary unit of replicated data. Our model is agnostic to the
unit of replication; it can be a bit, a sector, a file, a disk or an
entire storage site. We therefore attempt to develop a more abstract
model that can be interpreted in a more general, holistic fashion.
While still coarse-grained, our model is nonetheless helpful for
reasoning about the relative impact of a broader range of faults,
their detection times, their repair times, and their correlation. The
model helps point out what strategies are most likely to increase
reliability, and what data we need to measure in real systems to
resolve tradeoffs between these strategies.

We start with a simple, abstract definition of latent faults. We then
derive the mean-time-to-failure of mirrored data, in the face of both
immediately visible and latent faults. We extend this equation to
include the effect of correlated faults. Finally, we discuss the
implications of this equation for long-term storage.

\begin{figure}
\centerline{\includegraphics{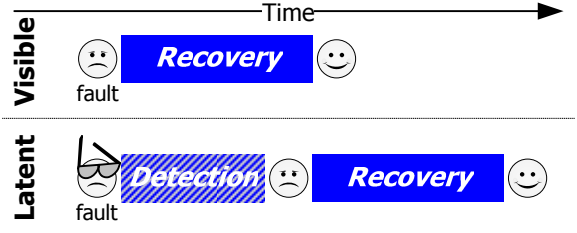}}
\caption{\small\em Types of replica faults.  
Time flows from left to right.  At the top, when a visible fault (sad
face) is detected, recovery begins immediately. At the end of
successful recovery, the fault has been corrected (smiley face).  At
the bottom, when a latent fault occurs (sad face in sunglasses),
nothing happens until the fault is detected.  Once the fault is
detected, as with visible faults, recovery takes place.}
\label{fig:FailureTypes}
\end{figure}

\subsection{Fault types}
\label{sec:analyticModel2:latent}

We assume that there are two types of faults: immediately visible and
latent, as shown in Figure~\ref{fig:FailureTypes}.
Visible faults are
those for which the time between their occurrence and detection is
negligible. Causes of such faults include entire-disk or
controller errors.  We denote the mean time to a visible fault by \MV
and the associated mean time to repair by \MRV.  Latent faults are
those for which the time between occurrence and detection is
significant.  Examples include misdirected writes, bit rot, unreadable
sectors, and data stored in obsolete formats. We denote the mean
time to a latent fault by \ML and mean time to repair by \MRL.  We
only consider latent faults that are detectable; hence, they have a
finite mean time between occurrence and detection, denoted by \MD.

\subsection{Assumptions}
\label{sec:analyticModel2:assumptions}

Our model is based on several assumptions. We build our model starting
from the simplest assumptions and increase its complexity as
needed. To start, with no additional information, the simplest
assumption we can make (as do others~\cite{Patterson1988short})
regarding the processes that generate faults (latent or visible) is
that they are memoryless. That is, the probability, $P(t)$, of a fault
occurring within time $t$, is independent of the past. This assumption
leads to the following exponential distribution

\begin{equation}
\label{eq:basic-exp}
P(t) = 1 - e^{t/ \MTTF}.
\end{equation}

\noindent where \MTTF is the mean time to the fault. For many parts of
our derivation, we consider the case where $t << \MTTF$, so the
following approximation holds

\begin{equation}
\label{eq:basic-approx}
1 - e^{t/ \MTTF} \approx 1 - (1 - \frac{t}{\MTTF}) = \frac{t}{\MTTF}.
\end{equation}

\noindent This approximation and similar ones below are used only to
simplify the expression for the exponential in the probability,
similar to Patterson {\em et al.}~\cite{Patterson1988}, so are not
fundamental to the model.

Initially, we also assume that all faults occur independently of one
another. Subsequently, we revise this assumption by introducing
correlated errors that are also exponentially distributed but with an
increased rate of occurrence. For simplicity, we model this increased
rate via a multiplicative correlation factor, which is assumed to be
the same for both latent and visible faults.

Furthermore, undetectable faults might also exist, but we ignore them
for this analysis. Their effect on the overall reliability will
dominate only if their rate of occurrence is significant. In such a
case, we must turn them into detectable faults, by developing a
detection mechanism for them, and correct them, by employing
redundancy~\cite{lampson1979}. Otherwise, they will remain the main
vulnerability for the stored data. 

\subsection{Reliability}
\label{sec:analyticModel2:reliability}

Mirrored data become irrecoverable when there are two successive
faults such that the copy fails before the initial fault can be
repaired, an event we call a double-fault. Since a double-fault leads
to data loss for mirrored replicas, for most of this section,
$\frac{1}{\MTTDL}$ is equal to the rate of double-fault failures,
where \MTTDL is the mean time to data loss. In this section, we derive
an expression for this quantity, which represents the reliability of
mirrored data, to understand how it is affected by visible, latent,
and correlated faults.  Note that the double-fault rate is meaningful
regardless of whether the faults causing those failures are detected
or even detectable. Thus, throughout this section, our reliability
analysis is from the perspective of the data rather than the
perspective of the user for whom errors can go unnoticed.

\begin{figure}
\centerline{\includegraphics{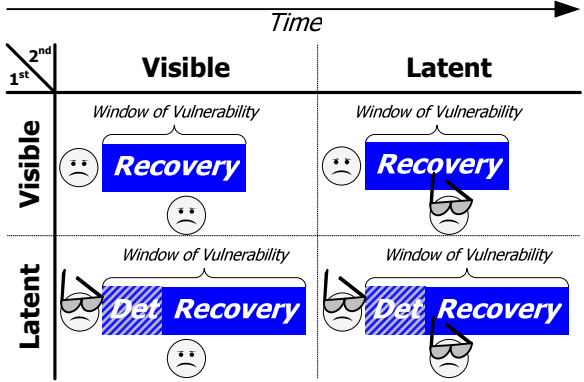}}
\caption{\small\em Combinations of
double faults resulting in data loss. The $y$ axis indicates the type of the
first fault (first sad face) and the $x$ axis indicates the type of the
second fault (second sad face). After the
first fault occurs, there is a window of vulnerability during which
the occurrence of a second fault will lead to data loss. After visible
faults, this window only consists of the recovery period.  After
latent faults, this window also includes the time to detect the
fault. If correlated, the second fault is more likely to occur within this window.}
\label{fig:DoubleFailures}
\end{figure}

To estimate \MTTDL, we first need to estimate the probability that a
second fault will occur while the first fault is still unrepaired. We
refer to this unrepaired period as the window of vulnerability
(WOV). Since there are two types of faults, we need to consider the
window of vulnerability after each type, as illustrated in
Figure~\ref{fig:DoubleFailures}.

First, consider the {\bf WOV after a visible fault}, $V_{1}$, which on
average is $\MRV$. During this WOV, both latent and visible faults can
occur. The probability that another visible fault, $V_{2}$ occurs is

\begin{equation}
\label{eq:pvv}
P(V_{2} | V_{1}) = \frac{ \MRV }{\MV}
\end{equation}

\noindent where $\MRV \ll \MV$. We obtain this result by using the
approximation in eqn~\ref{eq:basic-approx}.

The probability that another latent fault, $L_{2}$ occurs is

\begin{equation}
\label{eq:plv}
P(L_{2} | V_{1}) = \frac{ \MRV }{\ML}
\end{equation}

\noindent where $\MRV \ll \ML$. The difference between
$P(V_{2} | V_{1})$ and $P(L_{2} | V_{1})$
arises only from the different rates of fault occurrence.

Next, consider the {\bf WOV after a latent fault}, $L_{1}$, which on
average is $\MRL + \MD$. Again, during this WOV, both latent and
visible faults can occur, and the difference in the two probabilities
of occurrence is simply due to the different rates of occurrence for
visible and latent faults.

The probability that another visible fault, $V_{2}$ occurs is

\begin{equation}
\label{eq:pvl}
P(V_{2} | L_{1}) = \frac{ \MD + \MRL }{\MV}
\end{equation}

\noindent and the probability that another latent fault, $L_{2}$,
occurs is

\begin{equation}
\label{eq:pll}
P(L_{2} | L_{1}) = \frac{ \MD + \MRL }{\ML}
\end{equation}

\noindent As before, $\MRL + \MD \ll \MV$ and
$\MRL + \MD \ll \ML$.
Note that if $\MD$ becomes large,
the equations do not hold and the combined $P(V_{2} | L_{1}) + P(L_{2} | L_{1}) =
P(V_{2} \vee L_{2} | L_{1})$ approaches 1.

Next, we calculate the {\bf total double-fault failure rate} as follows

\begin{equation}
\label{eq:mttfdp}
\frac{1}{\MTTDL} = \frac{P(V_{2} | V_{1}) + P(L_{2} | V_{1})}{\MV} + \frac{P(V_{2} | L_{1}) + P(L_{2} | L_{1})}{\ML}
\end{equation}

\noindent where the first term on the right side counts the fraction
of the visible faults that result in double-failures, and the second
term counts the fraction of latent faults that result in double-failures.

To {\bf account for correlated faults}, we assume that the probability of
the second fault (conditioned on the occurrence of the first) is also
exponentially distributed, but with a faster rate parameter.
We introduce a multiplicative correlation factor
$\alpha < 1$ that reduces the mean time to the subsequent
fault once an initial fault occurs. In this case,
equations~\ref{eq:pvv}--\ref{eq:pll} are multiplied by $1 \over
\alpha$.  This is undoubtedly a vast simplification of how faults correlate in
practice. 
Modeling correlations accurately relies on modeling a particular
system instantiation and its component interactions, and often is
considered a black art.
We are at least not alone in using the
simplification~\cite{Corbett2004}.  An alternative would be to
introduce a distinct MTTF for correlated faults that is less than the
independent MTTF, as done for RAID by Chen {\em et al.} ~\cite{Chen1994}.

Combining the previous equations and accounting for correlated faults,
$\MTTDL$ becomes

\begin{equation}
\label{eq:mttfd}
\frac{\alpha \cdot {\ML}^{2}{\MV}^{2}}{(\MV + \ML)(\MRV \cdot \ML + (\MRL + \MD) \cdot \MV )}
\end{equation}

\subsection{Implications}
\label{sec:analyticModel2:implications}

To understand the implications of equation~\ref{eq:mttfd}, we
investigate its behavior at various operating ranges. We consider the
cases in which visible faults are much more frequent than latent
faults and vice versa. We also consider the case in which the WOV
after a latent fault occurs is long. We briefly discuss the implications in
each of these cases and touch upon reliability metrics for higher
levels of redundancy.

First, consider the case in which the visible fault rate dominates the
latent fault rate, $ \{ \MRL + \MD, \MRV \} \ll \MV \ll \ML $. Then,

\begin{equation}
\label{eq:mvdom}
\MTTDL \approx \frac{\alpha \cdot \MV^{2}}{\MRV}
\end{equation}

\noindent because $\MV + \ML \approx \ML$ and 
$\MRV \cdot \ML \gg (\MRL + \MD) \cdot \MV$. In this case, the effect
of latent faults is negligible, and thus the equation appropriately
resembles the original RAID reliability model~\cite{Patterson1988short}.

On the other hand, if the latent fault rate dominates the visible fault
rate, $ \{ \MRL + \MD, \MRV \} \ll \ML \ll \MV $, then

\begin{equation}
\label{eq:mldom}
\MTTDL \approx \frac{\alpha \cdot \ML^{2}}{\MRL + \MD}
\end{equation}

\noindent because $\MV + \ML \approx \MV$ and $\MRV \cdot \ML \ll (\MRL
+ \MD) \cdot \MV$. Equation~\ref{eq:mldom} indicates that if latent
faults are frequent, we must reduce $\MD$, as it can negate the
additional \ML factor of reliability, a result of replication.

Next, consider the case in which the visible fault rate dominates, $
\MRV \ll \MV \ll \ML $, but latent faults are frequent enough to be
non-negligible. If the window of vulnerability after a latent fault
occurs is long, then latent faults still play a role in increasing the
double-fault failure rate. The WOV can be long because either the
detection time for latent faults is long ($ \MD \approx \ML $), the
repair times are long ($ \MRL \approx \ML $), or both ($ \MRL + \MD
\approx \ML $). In this case, $ P( V_{2} \vee L_{2} | L_{1}) \approx 1
$ in equation~\ref{eq:mttfdp}, ensuring that a single latent fault is
extremely likely to lead to a double-fault failure. As a result, the
approximation

\begin{equation}
\label{eq:mdlong}
\MTTDL \approx \frac{\alpha \cdot \MV^{2}}{\MRV + \frac{\MV^2}{\ML}}
\end{equation}

\noindent holds when latent faults rates are non-negligible, i.e. $
\ML < \MV^{2}$.

These specializations of the model point to four important implications.
First, in equations~\ref{eq:mvdom} and~\ref{eq:mldom},
\MTTDL varies quadratically with both $\MV$ and $\ML$, and in
particular, with the minimum of $\MV$ and $\ML$. Thus, we must
consider occurrence rates for both fault types to improve overall
system reliability. We must be careful not to sacrifice one for the
other, which might happen in practice, since $\MV$ and $\ML$ can often be
anti-correlated depending upon hardware choice and detection strategy.

Second, equation~\ref{eq:mldom} indicates that if latent faults are
frequent, it is important to reduce their detection time, and not just
their repair time.  More specifically, consider a Seagate Cheetah disk
with a \MV of $1.4 \times 10^{6}$ hours, bandwidth of 300 MB/s, and
capacity of 146GB, leading to \MRV of 20 minutes.  Following Schwarz
{\em et al.}~\cite{Schwarz2004}), we assume that latent faults are
five times as likely as visible faults, resulting in an \ML of $2.8
\times 10^{5}$ hours.  Without scrubbing, we cannot justify the
approximations leading to equation~\ref{eq:mldom}. Therefore, applying
equation~\ref{eq:mttfdp} and substituting $P(V_{2} \vee L_{2} | L_{1})
\approx 1$, we achieve an $\MTTDL = 32.0$ years.  This gives a 79.0\%
probability of data loss in 50 years if we plug it into the
exponential distribution.

On the other hand, if we scrub a replica 3 times a year, as suggested by
Schwarz {\em et al.}, 
\MD is 1460 hours (which is half of the scrubbing
period). 
Then, applying equation~\ref{eq:mldom}, 
our reliability increases significantly. With no
correlated errors, $\MTTDL = 6128.7$ years, which gives a 0.8\% chance
of data loss in 50 years.  This implies that proactive searching for
latent faults is appropriate.  Other work agrees with this
conclusion~\cite{KariThesis,Prabhakaran2005,Schwarz2004}.

Third, in our model, correlation is a multiplicative factor and
affects the reliability regardless of the type of fault we have.
Continuing with our example above, assume $\alpha = 0.1$ as suggested
by Chen {\em et al.}~\cite{Chen1994}. Then, $\MTTDL = 612.9$ years,
which gives a 7.8\% chance of data loss in 50 years.

The above is a conservative assumption for $\alpha$ because this
correlation factor can vary by several orders of magnitude. To obtain
a reasonable lower bound on $\alpha$, consider that the correlated
mean-time-to-second-visible-fault can be an order of magnitude larger
than the recovery time; $\alpha \cdot \MV \geq 10 \cdot \MRV $, then
$\alpha \geq {10 \cdot \MRV \over \MV}$.  For example, a bug in the
firmware recovery code for a RAID controller might cause the
mean-time-to-second-fault to be not much more than the mean time to
recover. To obtain a specific lower-bound value, we assume the same
values as above for \MV and \MRV = \MRL, resulting in $1 \geq \alpha
\geq 2 \times 10^{-6}$, which gives a range of at least 5 orders of
magnitude.

Fourth, even when latent faults are infrequent,
equation~\ref{eq:mdlong} indicates that not attempting to detect
latent faults, or relying on very lengthy recovery procedures to fix
them, will leave the system vulnerable. For example, if $\ML = 1.4
\times 10^{7}$, \MV and \MRV remain the same, and $\alpha = 0.1$, then
$\MTTDL = 159.8$ years, leading to a 26.8\% probablility of data loss
in 50 years. In this case, because the system is negligent about
handling latent faults, the data are more susceptible to double-fault
failures from visible and latent faults following an initial
unrepaired latent fault.

\subsection{Replication and correlated faults}
\label{sec:analyticModel2:replication}

In this section, we show that additional replication does not offer
much additional reliability without independence.
To simplify our reliability analysis for higher degrees of
replication, we assume that we have instrumented the system so that
$\MD$ is negligible, and we assume that latent and visible faults
have similar rates and repair times.  We roughly estimate the
\MTTDL of a system with a degree of replication, $r$,
by extending the analysis along the same lines as for mirrored
data. We calculate the probability that $r-1$ successive, compounding
faults after an initial fault will leave the system with no integral
copy from which to recover.

To do so, we calculate the probability of $r-1$ successive faults
occurring within the vulnerability windows of each previous fault.
For simplicity, this analysis assumes that the vulnerability windows,
each of length $\MRV$, overlap exactly. In that case, the probability
that the $k$-th copy fails within the WOV of the previous $(k-1)$
failed copies is roughly $\frac{\MRV}{\alpha \cdot \MV}$.  Thus, the
probability that successive $r-1$ copies all fail within the WOV of
previously failed copies is the product of those $r-1$ probabilities,
$(\frac{\MRV}{\alpha \cdot \MV})^{r-1}$. Since the first fault occurs
with rate $1 \over \MV$, the overall mean-time-to-data-loss is

\begin{equation}
\MTTDL = \MV \cdot (\frac{\alpha \cdot \MV}{\MRV})^{r-1}
          = \frac{\alpha^{r-1} \cdot \MV^{r}}{\MRV^{r-1}}
\end{equation}

This equation shows that although increasing the level of replication, $r$,
geometrically increases $\MTTDL$, a high degree of correlated
errors ($\alpha \ll 1$) would also geometrically decrease $\MTTDL$,
thereby offsetting much or all of the gains from
additional replicas.

\section{Strategies}
\label{sec:strategies}

This simple model reveals a number of strategies for reducing the
probability of irrecoverable data loss.  While we generally describe them
in terms of the more familiar hardware and media faults,
they are also applicable to other kinds of faults.  For instance,
in addition to detecting faults due to media errors, scrubbing can
detect corruption and data loss due to attack.   As another example,
we can use a similar process
of cycling through the data, albeit at a reduced frequency, to detect 
data in endangered formats and convert to new formats  before we can
no longer interpret the old formats.

\begin{MYitemize}
\item Increase $\MV$ by, for example, using storage media less subject
to catastrophic data loss such as disk head crashes.
\item Increase $\ML$ by, for example, using storage media less subject
to data corruption, or formats less subject to obsolescence.
\item Reduce $\MD$ by, for example, auditing the data more
frequently to detect latent data faults,  as in RAID scrubbing.
\item Reduce $\MRL$ by, for example, automatically repairing latent
data faults rather than alerting an operator to do so.
\item Reduce $\MRV$ by, for example, providing hot spare drives so
that recovery can start immediately rather than once an operator has replaced
a drive.
\item Increase the number of replicas enough to survive more simultaneous
faults.
\item Increase $\alpha$ by increasing the independence of the replicas.
\end{MYitemize}

In the rest of this section we examine the practicality and costs of
some techniques for implementing these strategies.
We provide examples of systems using these techniques.

\subsection{Increase \MV or \ML}
\label{sec:strategies:MTTFV}

Based on Seagate's specifications, a 200GB consumer Barracuda
drive has a 7\% visible fault probability in a 5-year service
life~\cite{SeagateBarracuda}, whereas a
146GB enterprise Cheetah has a 3\% fault probability~\cite{SeagateCheetah}.
But the Cheetah costs about 14 times as much per
byte\footnote{\$8.20/GB versus \$0.57/GB.
Prices from TigerDirect.com 6/13/05.}.
The Barracuda has a quoted irrecoverable bit error rate of $10^{-14}$
and the Cheetah of $10^{-15}$. 
Even if the drives spend their 5 year life 99\% idle,
the Barracuda will suffer about 8 and the Cheetah about 6 irrecoverable
bit errors.
This 14-fold increase in cost between consumer and enterprise disk
drives yields approximately half the probability of in-service
fault and about $3/4$ the probability of irrecoverable bit
fault.  Thus, for long-term storage applications whose requirements for
latency and individual disk bandwidth are minimal,
the large incremental cost of enterprise drives
is hard to justify compared to
the smaller incremental cost of more (sufficiently independent)
replicas on consumer drives.

\subsection{Reduce \MD}
\label{sec:strategies:MD}

A less expensive approach to addressing latent faults due to media errors
is to detect the faults as soon as possible and repair them.
The only way to detect these faults is to \emph{audit} the replicas
by reading the data and either computing checksums or comparing against
other replicas.

Assuming (unrealistically) that the detection process
is perfect and the
latent faults occur randomly, $\MD$ will be half the interval between
audits, so the way to reduce it is to audit more frequently.
Another way to put this is that one can reduce
$\MD$ by devoting more disk read bandwidth to auditing
and less to reading the data; recent work suggests that in many systems
we can achieve a reasonable balance of auditing versus normal system
usage~\cite{Prabhakaran2005,Schwarz2004}.

On-line replicas such as disk copies have significant advantages
over off-line copies such as tape backups, for two
reasons.  First, the cost of auditing an off-line copy
includes the cost of retrieving
it from storage, mounting it in a reader, dismounting it and returning it
to storage.  This can be considerable, especially if the off-line copy
is in secure off-site storage.
Second, on-line media are designed to be accessed frequently and involve
no error-prone human handling.  They can thus be audited without the
audit process itself being as significant a cause of faults (up to some
limit, determined in part by the system strategy for powering components
on and off~\cite{Schwarz2004}).  Auditing off-line
copies, on the other hand, is a significant cause of highly correlated faults,
from the error-prone human handling of media~\cite{Reuters2005} to the
media degradation caused by the reading process~\cite{AMIA2003}.

The audit strategy is particularly important in the case of digital preservation
systems, where the probability that an individual data item will ever be
accessed by a user during a disk lifetime is vanishingly small.
The system cannot depend on user access to trigger fault detection
and recovery, because during the long time between accesses latent faults
will build up enough to swamp recovery mechanisms.  A system must therefore
aggressively audit its replicas to minimize $\MD$.  The
LOCKSS system is an example of such a system.

Note that relying on off-line replicas for security is not fool-proof.
Off-line storage may reduce the chances of some attacks, but it may
still be vulnerable to insider attacks.  Because it is harder to audit, the
damage due to such attacks may persist for longer.

\subsection{Reduce \MRL or \MRV}

Although the mean time to repair a latent media fault will normally be far less
than the mean time to detect it, and similarly the mean time to repair
a visible fault will be far less than the mean time to its occurrence,
reducing the mean time to repair is nevertheless important in reducing
the window during which the system is vulnerable to correlated faults.

Again in this case, on-line replicas have the major advantage that
repair times for media faults might be very short indeed, a few
media access times.
No human intervention is needed, and the process of repair is
in itself less likely to cause additional correlated faults.
Repairing from off-line media incurs the same high costs, long delays
and potential correlated faults as auditing off-line media.

\subsection{Increase replication}

Off-line media are the most common approach to increasing replication.
But the processes of auditing and recovering from faults using off-line
backup copies can be slow, expensive, and error-prone.

Some options for disk-based
replication strategies include replication within RAID
systems, across RAID systems, and across simple mirrored replicas.
Replication within RAID systems does not provide geographical or
administrative independence of the replicas.  If we opt for geographic and
administrative independence of replicas, it might be that the extra single-site
reliability provided by RAIDs is not worth the extra cost from a system-wide
perspective.  Further,
given the cost disparity between enterprise-grade drives
and consumer-grade drives (see \S~\ref{sec:introduction}),
adding more simple mirrored replicas in non-RAID configurations might
well be a cost-effective approach to increasing replication and thus
overall reliability.  OceanStore\cite{Kubiatowicz2000} is an example of using
a large number of replicas on cheap disks.

\subsection{Increase independence}
\label{sec:strategies:independence}

In just the six months of the Talagala study~\cite{Talagala1999}, many
correlated faults were observed, apparently caused by disks sharing power,
cooling, and SCSI controllers, and systems sharing network resources.
Our model suggests that in most cases, even with far lower rates of
correlated faults, increasing the independence of replicas is critical
to increasing
the reliability of long-term storage.

Long-term storage systems can reduce the probability of correlated
faults by striving for as much diversity as possible
in hardware, software, geographic location, and
administration, and by avoiding dependence on third-party components
and single organizations.
Examples include:

\begin{MYitemize}
\item \emph{Hardware}: Disks in an array often come from a single
manufacturing batch.  They thus have the same firmware, same hardware
and are the same age, and so are at the same point in the ``bathtub''
lifetime failure curve~\cite{Gibson1991thesis}.  However, the
increased cost that would be incurred by giving up supply chain
efficiencies of bulk purchase might make hardware diversity difficult.
Note, though, that replacing all components of a
large archival system at once is likely to be impossible.  If new
storage is added in ``rolling procurements'' over
time~\cite{Baker2005a}, then differences in storage technologies and
vendors over time naturally provides hardware heterogeneity.

\item \emph{Software}: Systems with the same software are vulnerable to epidemic
failure; Junqiera {\em et al.}\ have studied how the natural diversity of systems
on a campus can be used to reduce this vulnerability~\cite{Junqueira2005}.
However, the increased costs caused by encouraging such diversity,
in terms not merely of purchasing, but also of training and administration,
might again make this a difficult option for some organizations.
The British Library's system~\cite{Baker2005a}
is unusual in explicitly planning to develop
diversity of both hardware and software over time.
Nevertheless,
given the speed with which malware can find all networked systems sharing
a vulnerability,
increasing the diversity of both platform and application software
is an effective strategy for increasing $\alpha$.

\item \emph{Geographic location}:
Many systems using off-line backup store replicas off-site, despite
the additional storage and handling charges that implies.
Digital preservation systems, such as the British Library's,
establish each of their 
on-line
replicas in a different location, again despite the possible increased
operational costs of doing so.

\item \emph{Administration}:  Human error is a common cause of correlated
faults among replicas.
Again, the British Library's system is unusual in
ensuring that no single administrator will be able to affect more
than one replica.  This is probably more effective and more cost-effective
than attempts
to implement ``dual-key'' administration, in which more than one
administrator has to approve each potentially dangerous action.
In a crisis, shared pre-conceptions are likely to cause both
operators to make the same mistake~\cite{Reason1990short}.

\item \emph{Components}: System designs should avoid dependence
on third-party components that might
not themselves be preserved over time.  Determining all sources of such
dependence can be tricky, but some sources can be detected by running
systems in isolation to see what breaks.  For example, running a system
in a network without a domain name service or certificate authority can
determine whether the system is dependent on those services.  LOCKSS is
an example of a system built to be independent of the survival of
these services.

\item \emph{Organization}: Taking an end-to-end view of preservation systems, it
is also important to support their organizational independence.  For instance,
if the importance of a collection extends beyond
its current organization, then there must be an easy
and cost-effective ``exit strategy'' for the collection if the organization
ceases to exist.  As an example, bickering couples might not want the
survival of their
babies' photos to depend on a home IT system whose maintenance requires their
continued cohabitation.

\end{MYitemize}

In summary, the main techniques for increasing the reliability of long-term
storage are replication, independence of replicas, auditing replicas to
detect latent faults, and automated recovery to reduce repair times and costs.

\subsection{Tradeoffs}
\label{sec:strategies:tradeoffs}
Unfortunately, these strategies are not necessarily orthogonal, and some
can have adverse affects on reliability.  Here we consider the
effects of auditing and automated recovery. 
There are many other possible tradeoffs, such as the costs of
increased independence of administrative domains and diversity of
hardware and software that we do not cover here.

Auditing is necessary for detecting latent faults, but as previously
described, it increases the frequency of media access, which might increase
both visible and latent media errors and costs due to
increased consumption of power and other
system and administrative resources.  Previous
work~\cite{KariThesis,Prabhakaran2005,Schwarz2004} suggests it is possible
to achieve an appropriate balance that increases reliability considerably
while performing much of the audit work in background, even
opportunistically when legitimate data accesses require powering on the
corresponding system components.

Unfortunately, the audit process can itself introduce other channels for
data corruption.  An example would be attacking a distributed system through
the audit protocol itself~\cite{Maniatis2005short}, which therefore must be
designed as carefully as any other distributed protocol.

While automated recovery can reduce costs and
speed up recovery times, if buggy
or compromised by an attacker, it can itself
introduce latent faults. This can be dangerous because even
visible faults can now (though seemingly having been recovered) turn into
latent ones.

\subsection{Data gathering}
\label{sec:strategies:data}

Our simple model of reliability of replicated storage points out areas
where we are in great need of further data to validate the model and
to evaluate the potential utility of the reliability strategies
described in this paper. In particular, there is very little published
about the types and distribution of latent faults, both due to media
errors and also due to the other threats we describe. Moreover, the
correlations that result in latent faults are poorly understood. 

One desirable application of the model would be choosing, for
instance, between two levels of replication and audit.  Assume that
for disaster tolerance we have two geographically independent replica
systems.  Would it be better for each system to audit its storage
internally?  Or would it be better to audit between the two replicas?
Answering such a question requires understanding, at a minimum: the
$\MTTF$ of both visible and latent faults, the $\MD$ of different
audit strategies, the recovery strategies, the costs of replicating
information internally and geographically, and the costs of auditing
internally versus between sites.

To gather this information we can instrument existing systems. For a
start, we can log occurrences of visible faults, detection of latent
faults, and occurrences of data loss. One approach for detecting
latent faults is to cycle proactively through the storage logging
checksums of immutable objects. Ideally, these fault occurrences would
include timestamps and vertical information about the location of the
fault: block, sector, disk, file, application, etc. We would use such
information to determine the distributions and rates of the faults
that we consider, thereby testing the validity of our assumptions.
Additionally, we can log information about recovery procedures
performed (e.g., replacement of disks or recovery from tape), their
duration, and outcomes. We could use such data to measure mean
recovery times and, combined with the previous information, validate
the model itself. We could employ metadata about the configuration:
media, hardware, software, etc., to estimate costs. Finally, we could
log SMART data from disks, and external information such as
application workloads offered, processes spawned, file system and
network statistics, and administrator changes. We could mine such
information to identify and perform root cause analysis for correlated
faults.

Efforts are underway to gather some of this information for 
existing systems. For instance, several groups at UC Santa Cruz, HP
Labs and elsewhere have been processing just such failure data from
large archives such as the Internet Archive.  We need information,
though, from many different kinds of storage systems with different
replication architectures.

\section{Related Work}
\label{sec:related}

In this section we review related work,  showing how others address
problems,  such as correlated and latent faults,  that arise when
large amounts of data must remain unaltered yet accessible with low
latency at low cost.  We start with low-level approaches that focus
on single devices and RAID arrays,  then move up the stack.

Evidence of correlated faults comes from studies of disk farms.
Talagala~\cite{Talagala1999} logs every fault in a
large disk farm (of 368 disk drives) at UC Berkeley over six months
and shows significant correlations between them.  The study focuses
primarily on media (drive failures), power failures, and system
software/dependency failures.

Chen \emph{et al.}~\cite{Chen1994} explore the tradeoffs between performance
and reliability in RAID systems, noting that system
crashes (i.e., correlated faults) and uncorrectable bit errors (i.e., latent faults) can greatly
reduce the reliability predicted in the original RAID
paper~\cite{Patterson1988short}.  Kari's dissertation~\cite{KariThesis}
appears to be the first comprehensive analysis of latent disk failures.
His model also shows that they can greatly reduce the reliability of RAID
systems, and presents four scrubbing algorithms that adapt to disk
activity, using idle time to flush out latent errors.  In contrast, we
explore a broader space that includes application faults and distributed
replication.

Enterprise storage systems have recognized the need to address latent
and correlated faults.
Network Appliance's storage threat model includes two whole-disk
failures,  and a whole-disk failure with a latent fault discovered
during recovery (our $P( L_{2} | V_{1} )$). They employ row-diagonal parity
and suggest periodic scrubbing of disks to improve
reliability~\cite{Corbett2004}.
Schwarz \emph{et al.}~\cite{Schwarz2004} show that \emph{opportunistic} scrubbing,
piggy-backed on other disk activity, performs very well.  Like us,
they do not depend on the disk to detect a latent fault but actually
check the data. Our exploration also includes higher-layer failures.

Database vendors have implemented some application-level techniques to
detect corruption.  DB2's threat model includes a failure to
write a database page spanning multiple sectors atomically.
It sprinkles page consistency bits through each page, modifying and
checking them on each read and write. These bits detect some
other forms of corruption but only those affecting the consistency bits~\cite{Mohanpage1995}.
Tandem NonStop systems write checksums to disk
with the data, and compare them when data are read back~\cite{Bartlett2004}.

The IRON File System~\cite{Prabhakaran2005} is a file system whose
threat model includes latent faults and silent corruption of the disk.
It protects the file system's metadata (but not the data) using
checksums and in-disk replication.

An alternative to tightly-coupled replication such as RAID is more
loosely-coupled distributed replication.
In 1990 Saltzer suggested that digital archives need geographically
distributed replicas to survive natural disasters,  that they must
proactively migrate to new media to survive media obsolescence,
and that heavy-duty forward error correction could correct corruption
that can accumulate when data are rarely accessed~\cite{Saltzer1990}.
More recently,  these ideas inform the design of the British Library's
digital archive~\cite{Baker2005a}.

Many distributed peer-to-peer storage architectures have been proposed
to provide highly-available and persistent storage services, including
the Eternity Service~\cite{Anderson1996}, Intermemory~\cite{Chen1999},
CFS~\cite{Dabek2001}, OceanStore~\cite{Kubiatowicz2000},
PAST~\cite{rowstron01c}, and Tangler~\cite{mazieres01}.  Their threat
models vary, but include powerful adversaries (Eternity Service) and
multiple failures.  Some (OceanStore) use cryptographic sharing to
proliferate $n$ partial replicas from any $m<n$ of which they can
recover the data.  Others (PAST) replicate whole files.
Weatherspoon's~\cite{Weatherspoon2002a} model compares the reliability
of these approaches.  While the model itself does not include latent
errors, correlated errors, or others such as operational or human
errors, it takes into account the storage and bandwidth requirements of
each approach.  Later work~\cite{Weatherspoon2002b} identifies
correlations among replicas (e.g., geographic or administrative) and
informs the replication policy to reduce correlation effects.

Deep Store~\cite{You2005}
and the LOCKSS system~\cite{Maniatis2005short}
share the belief that preserving large amounts of data for long periods
affordably is a major design challenge.
Deep Store addresses the cost issues by eliminating redundancy,
LOCKSS by using a ``network appliance'' approach
to reducing system administration~\cite{Rosenthal2003b},
and large numbers of loosely coupled replicas on low-cost hardware.
Both recognize the threats of ``bit rot'' and format obsolescence
to long-term preservation.

\section{Conclusions and Future Work}
\label{sec:conclusion}

In this paper we motivate the need for long-term storage of digital
information and examine the threats to such information.
Using an extended reliability model that incorporates latent
faults, correlated faults, and the detection time of latent faults, we
reason about possible strategies for improving long-term reliability of these
systems.  The cost of these strategies is important, since limited budget is
one of the key threats to digital preservation.  We find that the most
important strategies are auditing to detect latent faults as soon as
possible, automating repair so that it is as fast, cheap, and as
reliable as possible, and increasing the independence of data replicas.

This is clearly a work in progress.  We do not yet have data to characterize
all the terms in our model.  The model thus points to several data collection
projects that would be very useful.  For instance, we need more data on the
mean time to different types of latent faults, the repair times
for faults once detected, and the levels of correlation of different kinds of
faults.  We are currently seeking out sources of these data and instrumenting
systems to gather more.

A digital preservation system should rarely suffer incidents that could
cause data loss. Thus the total experience base available to designers
of such systems will grow very slowly with time, making it difficult
to identify and fix the problems that will undoubtedly arise.
Past incidents indicate that it will in practice be very difficult to
accumulate this experience base.  The host organization's typical response to
an incident causing data loss is to cover it up.  A few details
become known via the grapevine, but the true story never becomes
part of the experience base.
A system similar to NASA's Aviation Safety Reporting System~\cite{ASRS}
should be established, through which operators of storage systems could
submit reports of incidents, even those not resulting in data loss,
for others to read in anonymized form and from which they can learn
how to improve the reliability of their own systems.

{\footnotesize \bibliographystyle{acm} \bibliography{../common/bibliography}}

\end{document}